\documentclass[aps,prd,nofootinbib]{revtex4-2}
\usepackage[colorlinks=true, pdfstartview=FitV, linkcolor=blue, citecolor=red, urlcolor=magenta]{hyperref}
\usepackage{graphicx}
\usepackage{latexsym}
\usepackage{amsmath}
\usepackage{amsfonts}
\usepackage{amssymb}
\usepackage{verbatim}

\newcommand{\be}{\begin{equation}}
\newcommand{\ee}{\end{equation}}
\newcommand{\bea}{\begin{eqnarray}}
\newcommand{\eea}{\end{eqnarray}}

\newcommand{\ben}{\begin{eqnarray}}
\newcommand{\een}{\end{eqnarray}}

\begin{document}

\title{Two-dimensional Lorentz-violating Casimir effect}


\author{$^{1,2}$K. E. L. de Farias}
\email{klecio.lima@uaf.ufcg.edu.br}

\author{$^{1}$M. A. Anacleto}
\email{anacleto@df.ufcg.edu.br}

\author{ $^{1,2}$F. A. Brito}
\email{fabrito@df.ufcg.edu.br}

\author{$^{1}$E. Passos}
\email{passos@df.ufcg.edu.br}

\author{$^{2}$ Herondy Mota}
\email{hmota@fisica.ufpb.br}


\affiliation{$^{1}$Departamento de F\'{\i}sica, Universidade Federal de Campina Grande,\\
Caixa Postal 10071, 58429-900, Campina Grande, Para\'{\i}ba, Brazil.}
\affiliation{$^{2}$Departamento de F\' isica, Universidade Federal da Para\' iba,\\  Caixa Postal 5008, Jo\~ ao Pessoa, Para\' iba, Brazil.}


\begin{abstract}
In this study, we consider the four-dimensional Maxwell electrodynamics extended with CPT-even Myers-Pospelov Lorentz-violating dimension-six operators to investigate the associated two-dimensional properties in the context of quantum vacuum fluctuation effects, namely, the Casimir effect. Upon projecting out the 4D theory down to a 2D theory we obtain analogs of these operators leading to a modified dispersion relation in a Lorentz invariance violation (LIV) scalar model equivalent to the electromagnetic theory. By making use of the modified dispersion relation, we derive exact analytic expressions for the Casimir energy and force induced by imposing Dirichlet boundary conditions on the scalar field. In the regime where the LIV parameter becomes very small, we recover known results for the Casimir energy and force plus correction terms due to the LIV.
\end{abstract}
\pacs{11.15.-q, 11.10.Kk} \maketitle


\section{Introduction}
\label{int}
The research of Lorentz invariance violation (LIV) in physics has garnered significant attention due to its potential to explain natural phenomena that the standard model of particle physics cannot account for, particularly in high-energy regimes near the Planck energy scale $(\sim 10^{19} \rm GeV)$. The LIV can be explored in various ways, such as by the notorious extension of the standard model \cite{Colladay:1998fq}, by the introduction of the aether field \cite{Carroll:2008pk}, by the Ho\v{r}ava-Lifshitz anisotropic theory \cite{Horava:2009uw}, and by the incorporation of high derivative operators \cite{Pospelov:2010mp}, which is the starting point of this study. Our focus is to elaborate an investigation of low-dimensional physics associated with LIV high derivative operators.

The study of several Lorentz-violating scenarios in the low-dimension field theory has been considered in Refs. \cite{Belich:2002vd, Bazeia:2005tb,Bazeia:2010yp,Passos:2008qh,SouzaDutra:2010szj,Casana:2011vh,Brito:2012yc,Passos:2016jur}. Among these works we can highlight as motivations for our investigation the one conducted in Ref.\cite{Belich:2002vd}, which considers the Lorentz-violating Chern-Simons model defined in 4D and performs a dimensional reduction to 3D generating a new model that mixes the gauge field, massless scalar field and an external vector which is used to verify the LIV effects in planar systems. Also, in Ref.\cite{Bazeia:2005tb}, the authors have conducted a 2D LIV study attributed to the scalar effective theory aiming to investigate the effects of LIV in the context of defect structures (see also Ref.\cite{SouzaDutra:2010szj}, which study a class of LIV travelling solitons systems). In addition, we highlight the study of Ref.\cite{Brito:2012yc}, which introduces the 2D LIV higher derivative term via radiative corrections and four-dimensional projection down to two dimensions. More specifically, this work considers the Myers-Pospelov
Lorentz-violating dimension-five operators to investigate 4D-2D projection, and then, apply in the defect scenarios (see also Ref.\cite{Passos:2016jur}). Here, in our investigation, we shall focus on 2D effects from the 4D theories with high derivative Lorentz-violating operators such as the 4D effective theories with Lorentz-violating
dimension-six operators.

We initially consider the gauge sector of the modified theory in order to derive the physics in two dimensions with analogues of these Lorentz-violating operators. As the main object of the dimensional reduction process, we shall consider the electrodynamics higher derivative operators of
dimension-six given as ${\mathcal L}_{4D} \sim \beta^{(6)} / M_{\mathrm{QG}}^2 F^{\mu \nu} u_\mu u^\sigma(u \cdot \partial)^2 \Tilde{F}_{\sigma \nu}$,
%
%
 with $\beta^{(6)}$ being a dimensionless parameters that control the LIV intensity over $M_{\mathrm{QG}}$ scale suppression, where $F_{\mu \nu}=\partial_\mu A_\nu-\partial_\nu A_\mu$, is the field-strength tensor \cite{Myers:2003fd}. The main property of this term is that it modifies the dispersion relation with a quadratic term to the energy content of the system \cite{Liberati:2009pf}. Thus, we take advantage of this procedure to obtain a 2D theory with scalar fields coupled to dimensional-four operators. Our goal is to investigate new LIV effects on physical observables associated with 2D systems, as for instance, the Casimir energy and force.


The detection of a LIV event implies the existence of new physics to explore, complementing the current standard model of particle physics. However, experiments aimed at detecting and measuring LIV are inherently challenging. The increasing sensitivity of experiments designed to detect Casimir forces offers a novel approach to study this phenomenon. The Casimir effect is a natural occurrence that arises when two parallel plates are placed close to each other (typically within micrometers). This arrangement limits the frequencies of vacuum fluctuations between the plates, resulting in a difference in force between the interior and exterior of the plates. First discovered by H. Casimir in 1948 \cite{Casimir:1948dh}, an attempt to detect it was initially made by Sparnaay in 1958 \cite{Sparnaay:1958wg}. However, recent advancements have allowed for high-precision detection by various experiments \cite{Lamoreaux:1996wh,Mohideen:1998iz,Bressi:2002fr}, highlighting the evolving sensitivity in detecting this force.

With the evolving sensitivity in detecting Casimir forces, the idea of a Casimir theory that violates Lorentz symmetry becomes increasingly plausible and extensively explored in the literature \cite{MoralesUlion:2015tve,ADantas:2023wxd,Erdas:2023wzy}. In such models, the LIV parameter introduces corrections to the Casimir energy. However, these proposals often result in expressions without exact analytic solutions, necessitating the use of approximations to compute the Casimir energy and force. In our approach, however, we provide exact analytic solutions without resorting to approximations. The significance of obtaining a closed-form solution lies in the ability to fully understand the behaviour of the system and explore its limits, thereby elucidating the physical constraints of the parameters involved.

The present work is structured as follows: In Section \ref{sec2}, we develop the dimensional projection of a 4D theory into a 2D
one for the gauge sector of an extended higher derivative electrodynamics. The Section \ref{sec3} addresses the derivation of the modified equation of motion and provides insights into the modified dispersion relation. Moving on to Section \ref{sec4}, we apply Dirichlet boundary conditions on the free solution of the modified equation of motion and use the corresponding modified dispersion relation to obtain exact analytic expressions for the 2D Casimir energy, including its approximation for the regime of small values for the LIV parameter, demonstrating the recovery of known results found in literature. Section \ref{sec5} focuses on deriving the analytic expression for the Casimir force, along with its approximation for small values of the LIV parameter, again recovering the standard result. Finally, in Section \ref{sec6}, we offer our conclusions based on the results obtained. Throughout this paper, we utilize natural units with $\hslash=c=1$.

\section{4D-2D DIMENSIONAL PROJECTION}
\label{sec2}


In this initial point, we focus on the dimensional projection out of the four-dimensional gauge field components of a $4 \mathrm{D}$ theory into a 2D theory. And, as previously mentioned, we consider the Maxwell electrodynamics plus a CPT-even higher derivative six-operator \cite{Liberati:2009pf}. The Lagrangian density for this configuration of the system is then given by
\bea
\mathcal{L}_{4D}=-\frac{1}{4} F_{\mu \nu} F^{\mu \nu}
-\frac{\bar{g}}{2}(u \cdot \partial) F^{\mu \nu} u_\mu u^\sigma(u \cdot \partial) \Tilde{F}_{\sigma \nu},
\label{eq1}
\eea
where $\bar{g}=\beta^{(6)} / M_{\mathrm{QG}}^2$ is the LIV parameter with dimension of length squared. Notice that the other quantities in Eq.(\ref{eq1}) have already been introduced in Sec.\ref{int}.

Our simple 4D-2D dimensional projection will follow the approach of Ref.\cite{Brito:2012yc}, which implies to say that $\partial_{1}=\partial_{2}=0$ when acting on any field (therefore, $\left.F_{a b}=0\right)$ being $a, b=1,2$. In this way,
the Maxwell term assumes the form
\begin{align}
-\frac{1}{4} F_{\mu \nu} F^{\mu \nu} & =-\frac{1}{4}\left(F_{i j} F^{i j}+2 \partial_0 A_a \partial^0 A^a+2 \partial_3 A_a \partial^3 A^a\right)\nonumber \\
& =-\frac{1}{4} F_{i j} F^{i j}+\frac{1}{2} \partial_i \phi \partial^i \phi+\frac{1}{2} \partial_i \chi \partial^i \chi,
\label{e2}
\end{align}
where $i, j=0,3$. We will work with the metric signature $(+,-,-,-)$, where $A_1=\phi$ and $A_2=\chi$.
Let us now discuss the CPT-even extension. Under the same conditions described above we obtain
\begin{align}
-\frac{\bar{g}}{2}(u \cdot \partial) F^{\mu \nu} u_\mu u^\sigma(u \cdot \partial) \Tilde{F}_{\sigma \nu}= & -\frac{\bar{g}}{2}\left(F_{i j} u^i u_k\left(u_l \partial^l\right)^2 F^{k j}+\left(\partial_i A_a\right) u^i u_j\left(u_l \partial^l\right)^2\left(\partial^j A^a\right)\right)\nonumber \\
= & -\frac{\bar{g}}{2} F_{i j} u^i u_k\left(u_l \partial^l\right)^2 F^{k j}+\frac{\bar{g}}{2}\left(\partial_i \phi\right) u^i u_j\left(u_l \partial^l\right)^2\left(\partial^j \phi\right)+ \nonumber\\
& \frac{\bar{g}}{2}\left(\partial_i \chi\right) u^i u_j\left(u_l \partial^l\right)^2\left(\partial^j \chi\right).
\label{e3}
\end{align}
Therefore, from a simple method of 4D-2D dimensional projection, we can construct other LIV effective models in two-dimensions. Among them, a LIV scalar model equivalent to the electromagnetic theory given by Eq.(\ref{eq1}) can be similar to
\begin{equation}
    \mathcal{L}_{2 D}=\frac{1}{2} \partial_{i} \phi \partial^{i} \phi+\frac{\bar{g}}{2}\left(\partial_i \phi\right) u^i u_j\left(u_l \partial^l\right)^2\left(\partial^j \phi\right) + \cdot\cdot\cdot,
    \label{e4}
\end{equation}
where $(\cdot\cdot\cdot)$ means terms associate to $\chi-$field. This resulting model is made out of two-dimensional fields living in two dimensions. In two dimensions, the new term corresponds to a four-dimensional operator Lorentz-violating. We shall refer to it as the 2D CPT-even higher-derivative scalar model.

\section{A Brief Analysis of the Model}
\label{sec3}

Considering the result given by Eq.(\ref{e4}), we write the following $(1+1)D$ massless Lagrangian density:
\begin{equation}
   \mathcal{L}_\phi=\frac{1}{2} \partial_\mu \phi \partial^\mu \phi+\frac{\bar{g}}{2}\left(\partial_\mu \phi\right) u^\mu u_\nu\left(u_\alpha \partial^\alpha\right)^2\left(\partial^\nu \phi\right),
   \label{e5}
\end{equation}
where $\mu, \nu = 0, 1$ with the constant parameter, $u_\mu=\left(u_0, u_1\right)$, controlling the LIV effect in the theory. Notice that we are working in $2 \mathrm{D}$ such that the metric signature is given as $(+,-)$.
The higher-derivative equation of motion associated with the Lagrangian, $\mathcal{L}_\phi$, can be derived from
\begin{equation}
    \frac{\partial \mathcal{L}_\phi}{\partial \phi}-\partial_\lambda\left(\frac{\partial \mathcal{L}_\phi}{\partial\left(\partial_\lambda \phi\right)}\right)+\partial_\lambda \partial_\rho\left(\frac{\partial \mathcal{L}_\phi}{\partial\left(\partial_\lambda \partial_\rho \phi\right)}\right)-\partial_\lambda \partial_\rho \partial_\sigma\left(\frac{\partial \mathcal{L}_\phi}{\partial\left(\partial_\lambda \partial_\rho \partial_\sigma \phi\right)}\right)+\cdots=0.
    \label{e6}
\end{equation}
Consequently, we obtain a modified equation of motion as follows
\begin{equation}
    \square \phi+\bar{g} u^\alpha u^\beta u^\mu u^\nu\left(\partial_\alpha \partial_\beta \partial_\mu \partial_\nu \phi\right)=0.
    \label{e7}
\end{equation}
This equation offers us the following set of equations:
\begin{subequations}
    \begin{equation}
     \ddot{\phi}-\phi^{\prime \prime}+\bar{g} \ddddot{\phi}=0, \quad {\rm time - like\; approach}: u_\mu=\left(1, 0\right),
     \label{e8.1}
    \end{equation}
    \begin{equation}
    \ddot{\phi}-\phi^{\prime \prime}+\bar{g} \phi^{\prime \prime \prime \prime}=0, \quad {\rm space - like\; approach}: u_{\mu}=\left(0, 1\right),
    \label{e8.2}
    \end{equation}
\end{subequations}
where dots and primes stand for derivatives with respect to time $t$ and position $x$, respectively.

The solution to Eqs. \eqref{e8.1} and \eqref{e8.2} is of the plane wave type, that is, $\phi\propto e^{-iEt+ikx}$, with $k$ being the momentum in the $x$-direction. Consequently, we can derive the modified dispersion relation associated with this solution. The equation of motion associated with the time-like case, Eq.(\ref{e8.1}), along with the plane wave solution, provide the following energy-momentum equation:

\begin{equation}
    E^2-k_x^2+\bar{g} E^4=0.
    \label{rd02}
\end{equation}
We get a set of the following solutions
\begin{align}
 E=& \pm \frac{1}{\sqrt{2}} \sqrt{\frac{-1-\sqrt{1+4 \bar{g} k_x^2}}{\bar{g}}}, \label{rd02a}\\
 E=& \pm \frac{1}{\sqrt{2}} \sqrt{\frac{-1+\sqrt{1+4 \bar{g} k_x^2}}{\bar{g}}}.\label{rd02b}
\end{align}

\begin{figure}[h]
    \centering
    \includegraphics[scale=0.5]{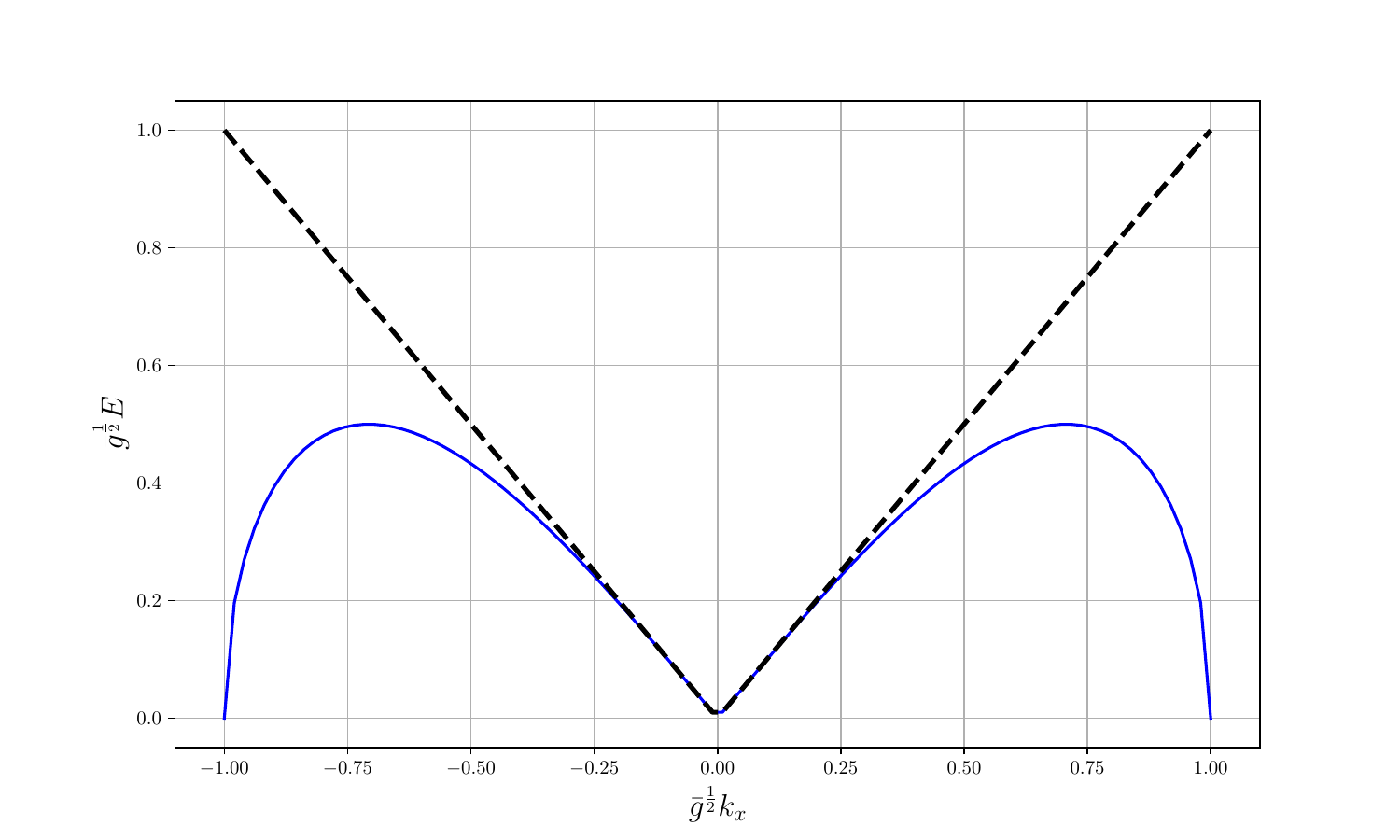}
    \caption{The behaviour of the dispersion relation in the time-like approach, Eq. \eqref{rd01a}. The black dashed line represents the light cone and the blue solid line shows the energy $E$. }
    \label{fig01}
\end{figure}

Note that the first set of equations yields imaginary solutions independent of the value of $k_x$, and these solutions should be discarded. However, the second set of equations leads us to real solutions. In the regime of $4\bar{g}k_x^2\ll1$, we have the following expansion:
$\sqrt{1+4 \bar{g} k_x^2}=1+\frac{4 \bar{g} k_x^2}{2}-\frac{(4 \bar{g} k_x^2)^2}{8}$. Therefore, Eq. \eqref{rd02b} is reduced to
\bea\label{rd01a}
E(k_{x})= |k_{x}|\sqrt{1 - \bar{g}  k_{x}^{2}}.
\eea
This solution is similar to the dispersion relation associated with 4D theory (see Ref.\cite{Liberati:2009pf}) and provides evidence for the LIV. Moreover, we can also note that Eq. \eqref{rd01a} lead us to the loss of unitarity and instabilities (we plot the energy solution, Eq.(\ref{rd01a}), in Fig.\ref{fig01}).

\begin{figure}[h]
    \centering
    \includegraphics[scale=0.5]{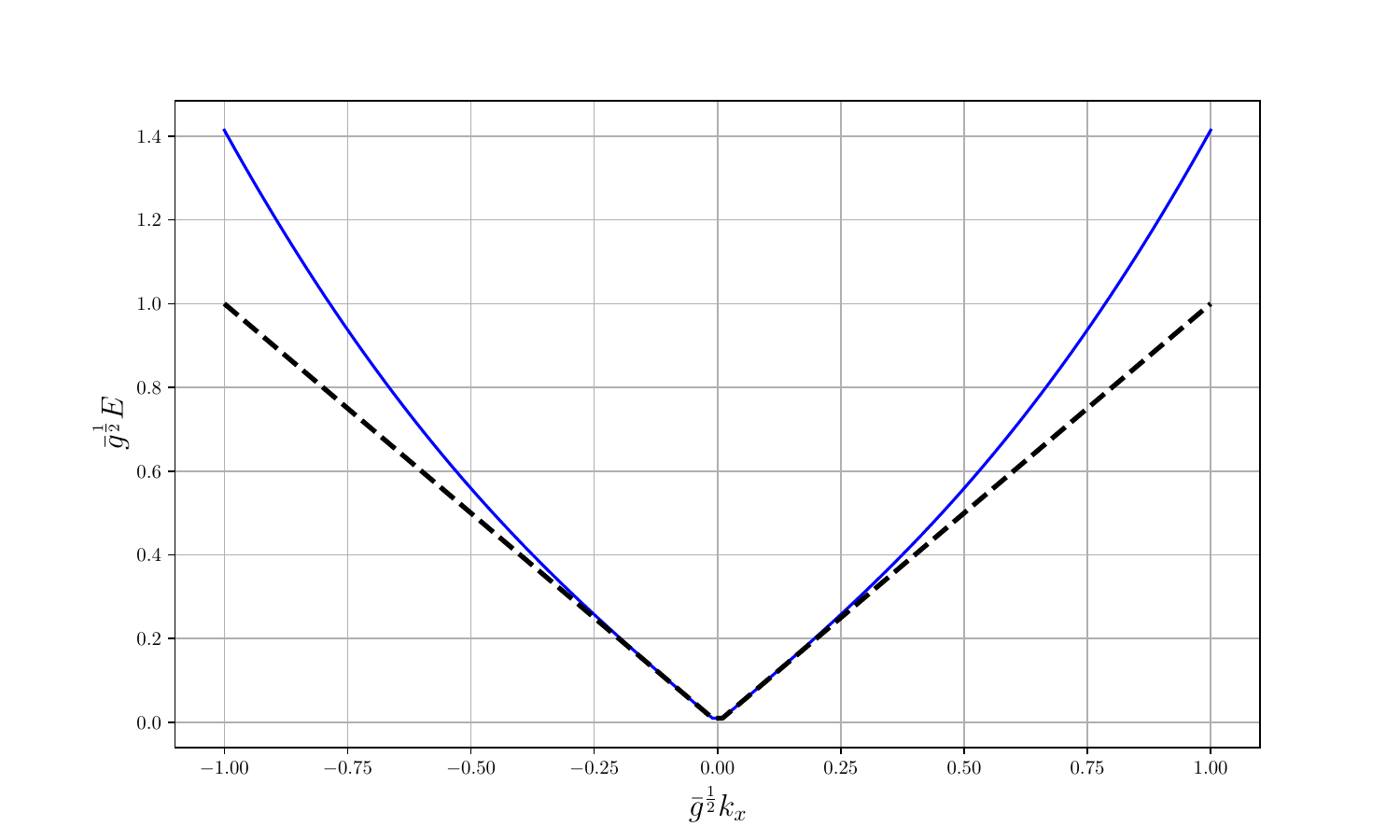}
    \caption{The behaviour of the dispersion relation in the space-like approach, Eq. \eqref{rd01bb}. The black dashed line represents the light cone and the blue solid line shows the energy $E$. }
    \label{fig02}
\end{figure}

On the other hand, the space-like case dispersion relation associated with the motion of equation, Eq.(\ref{e8.2}), is given as
\bea\label{rd01b}
E^{2} -k^{2}_{x} - \bar{g}  k^{4}_{x}=0.
\eea
This leads us to the following energy solution
\bea\label{rd01bb}
E(k_{x})= |k_{x}|\sqrt{1 + \bar{g}  k_{x}^{2}}.
\eea
Differently from the time-like case, this solution has no imaginary contribution, and then, with no loss of stability for approximation of higher momenta (see its behaviour in  Fig.\ref{fig02}).
%

At this point, we present a simple analysis of group and phase velocities. Considering only the dispersion relation, given by Eq.(\ref{rd01bb}), leads to group and phase velocities, respectively,
\begin{subequations}
    \bea\label{velg}
v_{g}(k_{x}) = \frac{(1 + 2 \bar{g} k_{x}^{2})}{\sqrt{1 + \bar{g}k_{x}^{2}}},
    \eea
    \bea\label{velp}
v_{p}(k_{x}) = \sqrt{1 + \bar{g}k_{x}^{2}}.
    \eea
\end{subequations}
Note that the phase and group velocities above are related through Rayleigh’s formula: $v_{p}/v_{g} = 1 - \big(k/v_{g}\big)\big(dv_{p}/d k_{x}\big)$. Hence, from Eqs.(\ref{velg})--(\ref{velp}), we find
\bea\label{rfr}
\frac{v_{p} - v_{g}}{v_{g}} &=&-\frac{\bar{g}k_{x}^{2}}{1 + 2\bar{g}k_{x}^{2}}\nonumber\\
&\approx&-\bar{g}k_{x}^{2} + {\cal O}(\bar{g}k_{x}^{2}),
\eea
for the condition $\bar{g}|\vec{k}|^{2} \ll 1$. Note for $\bar{g}>0$ implies that $v_{g} > v_{p}$ which means an anomalous medium associated with the superluminal propagation (from an influence of anisotropy, i.e., that refers to the medium in which the properties are different in some directions, namely crystalline solids in nature.) Moreover, when the group velocity of a light pulse exceeds the speed of light in a vacuum, it implies that the dispersion medium is sufficiently steep \cite{brown2008anomalous}.


\section{Casimir Energy}
\label{sec4}

As shown earlier, the current model presents two alternatives for expressing energy: the time-like approach, given by \eqref{e8.1}, as well as the space-like approach, given by \eqref{e8.2}. While both can be considered for analyzing the system's energy, the timelike approach introduces inconsistencies to the energy when examining the dispersion relation, as explained in the last section. Consequently, we exclusively use the spacelike approach to derive the energy density.

By examining the dispersion relation obtained for the spacelike approach, we can determine the vacuum energy fluctuation, i.e., the Casimir energy. In order to do that we aim to consider a configuration constituting of two parallel plates, similar to the original Casimir system, which leads us to an energy density depending on the volume between the plates. However, due to the dimensional reduction, the effect of the plates is in fact codified in two points along the line in the $x$-direction. Hence, in our effective $(1+1)D$ system, by applying Dirichlet boundary conditions on the free solution of the equation of motion discussed in the previous section, the dispersion relation $E^2 = k_x^2 + \bar{g}k_x^4$ now with discretized momentum in the $x$-direction becomes
\begin{align}
E=&\frac{1}{2}\sum_{\textcolor{blue}{n=1}}^{\infty} \sqrt{\left(\frac{n \pi}{a}\right)^2+\bar{g}\left(\frac{n \pi}{a}\right)^4},
\label{c1}
\end{align}
where $a$ is the distance between the boundary condition points. It is clear that the sum in \eqref{c1} yields a divergent result. However, we can employ the Abel-Plana formula to obtain the finite part of this sum, and discard the infinite continuum contribution by means of the minimum subtraction scheme renormalization.

For our case, the Abel-Plana formula is expressed as follows \cite{saharian2007generalized}:
\begin{equation}
    \sum_{n=0}^{\infty} f(n)=\int_0^{\infty} f(z) d z+\frac{1}{2} f(0)+i \int \frac{f(i z)-f(-i z)}{e^{2 \pi z}-1} d z,
    \label{abel}
\end{equation}
where, from Eq. \eqref{c1}, we have defined
\begin{align}
f(n)=&\frac{1}{2} \sqrt{\left(\frac{n \pi}{a}\right)^2+\bar{g}\left(\frac{n \pi}{a}\right)^4}.
\label{func}
\end{align}
Upon using the above equation, the first integral in the r.h.s. of Eq. \eqref{abel} gives us the divergent free Minkowski contribution and, consequently, should be discarded by means of the minimum subtraction scheme. Note also that $f(0)=0$.

Hence, the finite part of the sum of \eqref{func} is provided by the third term in the r.h.s. of the Abel-Plana formula \eqref{abel}, that is,
\begin{align}
E_0^{\rm ren } & =\frac{i}{2} \int_0^{\infty} \frac{f(i z)-f(-i z)}{e^{2 \pi z}-1} d z \nonumber\\
&=i \frac{a}{2 \pi} \int_0^{\infty} \frac{d u}{e^{2 a u}-1}\left[\sqrt{i^2 u^2+\bar{g} u^4}-\sqrt{(-i)^2 u^2+\bar{g} u^4}\right],
\label{c2}
\end{align}
where we have made the change of variable $u=\frac{z \pi}{a}$. In order to solve the integral above one should note that for $u>\bar{g}^{-\frac{1}{2}}$ we obtain a null result since
\begin{equation}
f(i z)-f(-i z)=0.
 \label{c3}
\end{equation}
However, for $u<\bar{g}^{-\frac{1}{2}}$, we have
\begin{equation}
f(i z)-f(-i z)=2i\sqrt{u^2 - \bar{g}u^4}.
 \label{c3}
\end{equation}
As a result, Eq. \eqref{c2} becomes
\begin{equation}
E_0^{\rm ren}=-\frac{a}{\pi\bar{g}}\int_0^{1} \frac{\sqrt{y^2-y^4}}{e^{2xy}-1}dy,
\label{c4}
\end{equation}
 where we have made the new change of variable $y=\bar{g}^{\frac{1}{2}}u$ and defined the dimensionless parameter $x=a\bar{g}^{-\frac{1}{2}}$. By using the identity
\begin{equation}
\frac{1}{e^{2xy}-1}=\sum_{k=1}^{\infty}e^{-2xyk},
\label{identity}
\end{equation}
the integral \eqref{c4} can be performed to provide \cite{gradshteyn2014table}
\begin{align}
E_0^{\rm ren}=-\frac{a}{\pi\bar{g}} \sum_{k=1}^{\infty}\left[\frac{2 \mu_{k}-3 \pi I_2(\mu_{k})+3 \pi L_2(\mu_k)}{6 \mu_{k}}\right],
\label{c5}
\end{align}
where $\mu_{k}=2xk$, $I_n(x)$ is the modified Bessel function of the first kind and $L_n(x)$ is the modified Struve function. Alternatively, the renormalized vacuum energy can be rewritten as
\begin{equation}
E^{\rm ren}_0=\frac{x^2}{6 \pi a}-\frac{x}{4 a} \sum_{k=1}^{\infty} \frac{1}{k}\left[L_2(2 kx)-I_2(2 kx)\right],
\label{c5.1}
\end{equation}
where we have used the Riemann zeta function $\zeta(s)$ calculated at zero argument, i.e., $\zeta(0)=-\frac{1}{2}$, to arrive at the first term of the r.h.s. of \eqref{c5.1}. The expressions in Eqs. \eqref{c4} and \eqref{c5.1} are equally equivalent, however, as we shall see below, the former is the best one for numerical computations while the latter is more convenient to analyze the regime of small values of the LIV parameter $\bar{g}$ or, equivalently, $x=a\bar{g}^{-\frac{1}{2}}\gg 1$.  This is an important regime to study since it allows us to verify the consistence of our results by showing the recovery of the standard result, i.e., the result in the absence of LIV.

\begin{figure}[h]
    \centering
    \includegraphics[scale=0.5]{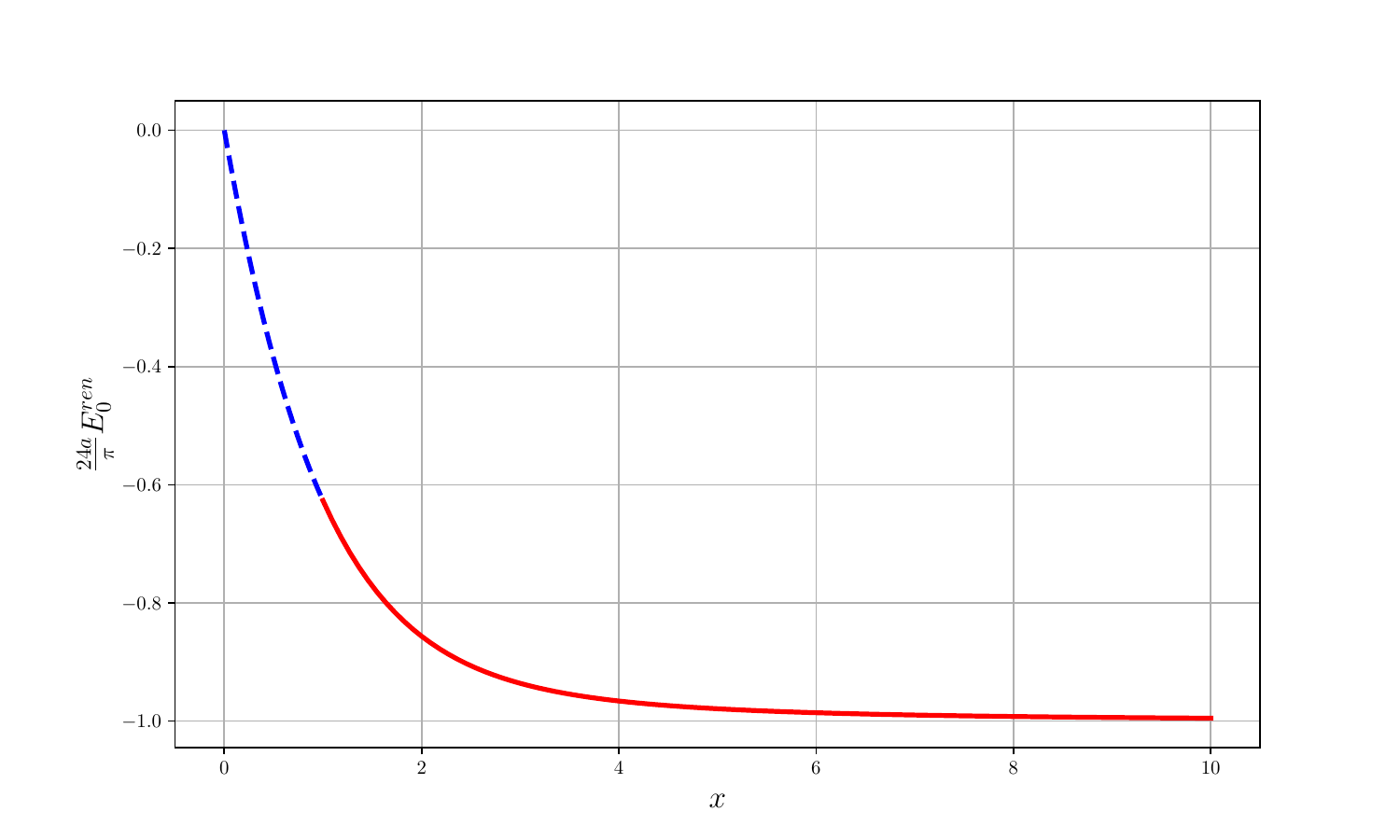}
    \caption{The behaviour of Eqs. \eqref{c4} and \eqref{c5.1}, expressed dimensionless as $\frac{24\pi a}{\pi}E^{\rm ren}_0$, in terms of $x=a\bar{g}^{-\frac{1}{2}}$.}
    \label{fig1}
\end{figure}

The behaviour of the renormalized energy is described in Fig.\ref{fig1}, which we have plotted by using Eq. \eqref{c4}.  The plot shows the curve traced by the energy, per unit of $\frac{\pi}{24a}$, in terms of $x=a\bar{g}^{-\frac{1}{2}}$. We can see that in the limit of small LIV parameter, i.e., $x=a\bar{g}^{-\frac{1}{2}}\gg 1$, the curve (red solid line) tends to the known result for the vacuum energy in $(1+1)D$ \cite{ambjorn1983properties}, while in the opposite limit it tends to zero (blue dashed line). However, from the experimental point of view, the limit $x\ll 1$ does not make sense since it would indicate a strong signal of LIV. Thus, the part of the curve that does make sense is indicated in red solid line.

We can analytically investigate the limit where the Lorentz violating parameter is small, $x\gg 1$, as previously mentioned. In order to obtain this limit, we need to consider the approximation below, for large arguments, of the modified Struve function  \cite{abramowitz1988handbook}:
\begin{align}
& L_\nu(z)-I_{-\nu}(z) \simeq \frac{1}{\pi} \sum_{j=0}^{\infty} \frac{(-1)^{j+1} \Gamma(j+1 / 2)}{\Gamma(\nu+1 / 2-j)}\left(\frac{2}{z}\right)^{2 j-\nu+1},
\label{c6.0}
\end{align}
where $\Gamma(s)$ is the Gamma function. Having in mind that $I_{-n}(z)=I_n(z)$ ($n \in \mathbb{Z}$), the use of the above approximation in Eq. \eqref{c5.1} provides
\begin{equation}
E^{\text {ren }}_0=\frac{x^2}{6\pi a}-\frac{x}{4a\pi}\sum_{j=0}^{\infty} \frac{(-1)^{j+1} \Gamma(j+1 / 2)}{\Gamma(5/2-j)}\frac{\zeta(2j)}{x^{2j-1}}.
\label{c7}
\end{equation}
%
In order to have a better understanding of what the above approximation gives, let us consider it up to order $\bar{g}^3$, that is,
\begin{equation}
\begin{aligned}
E^{\text {ren }}_0=-\frac{\pi}{24a}+\frac{\bar{g} \pi^3}{480 a^3}+\frac{\bar{g}^2 \pi^5}{4032 a^5} +\frac{\bar{g}^3 \pi^7}{7680 a^7} + \mathcal{O}(\bar{g}^4).
\label{c8}
\end{aligned}
\end{equation}
We can observe that the first term in the r.h.s. of the above expression is the known result for the vacuum energy in $(1+1)D$, assuming Lorentz invariance is preserved \cite{ambjorn1983properties}, as indicated in the plot of Fig.\ref{fig1}. The other terms are corrections to the latter due to the high derivative operator, which causes LIV, present in the Lagrangian \eqref{e5}. Therefore, the presence of the LIV parameter allows us to access correction terms of high orders. These correction terms, as shown in Eq. \eqref{c8} and confirmed by the plot in Fig.\ref{fig1}, makes the vacuum energy to increase. Again, the effect of the corrections term, as it has happened also in the energy's case, is to increase the vacuum force making it less attractive.

A similar energy correction to the approximation result \eqref{c8} is obtained in Ref. \cite{Moharramipour:2023dmh}, where the authors find a correction energy dependent on the inverse of the distance in a one-dimensional Ising antiferromagnetic model. Moreover, other applications considering the Casimir effect in a one-dimensional system can be found in the literature. Examples of such applications include studies on one-dimensional quantum liquids \cite{recati2005casimir}, Planck radiation \cite{Lynch:2022rqx}, and Bose gases \cite{Reichert:2019adr}.

\section{Casimir force}
\label{sec5}

\begin{figure}[h]
    \centering
    \includegraphics[scale=0.5]{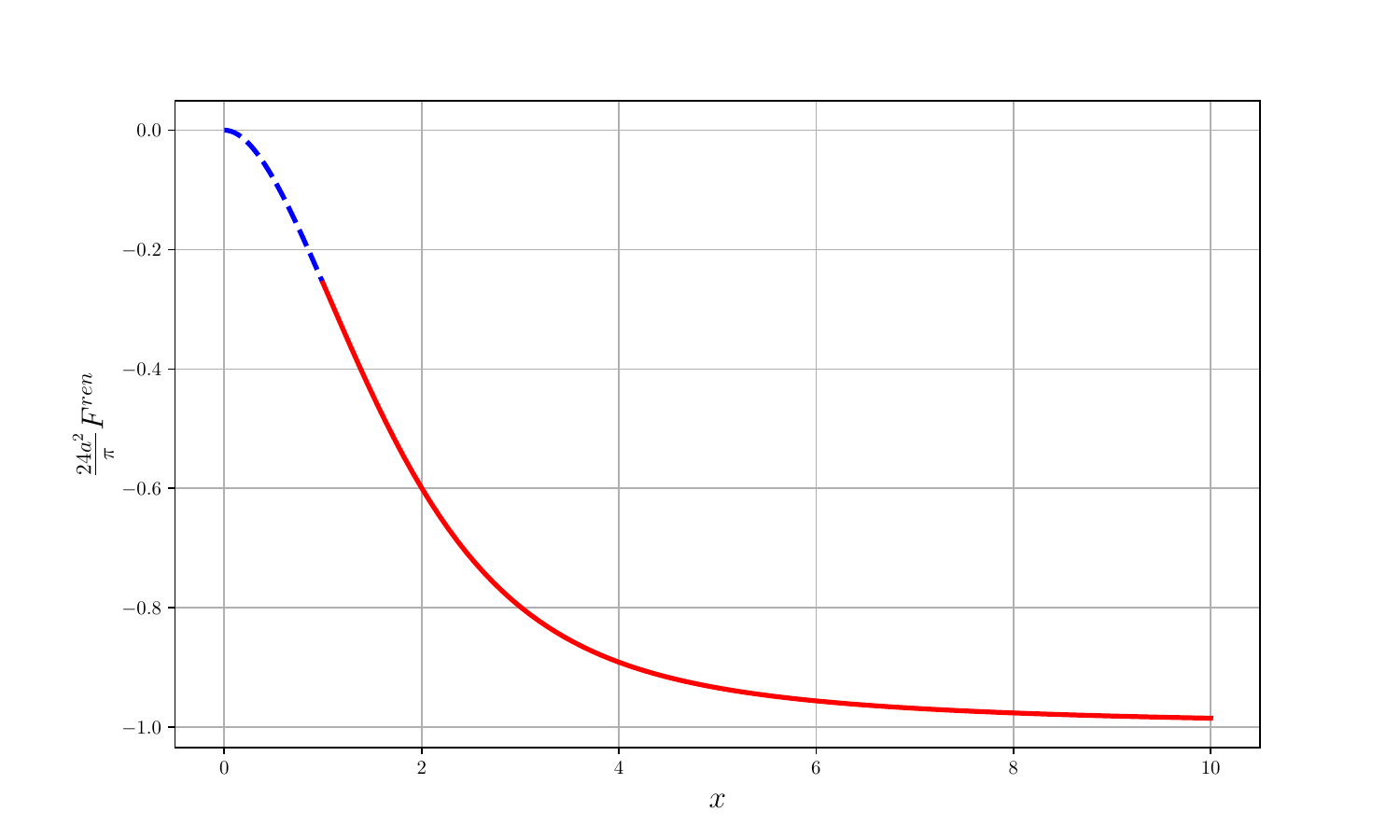}
    \caption{The behaviour of the vacuum force \eqref{c9}-\eqref{c9.1}, expressed dimensionless as $\frac{24\pi a^2}{\pi}F^{\rm ren}$, in terms of $x=a\bar{g}^{-\frac{1}{2}}$.}
    \label{fig3}
\end{figure}

The Casimir force arising from the Dirichlet boundary condition is obtained by taking the derivative of the energy with respect to the distance $a$. Hence, starting from \eqref{c4}, which is an analytical solution without any approximation, we obtain:
\begin{align}
F^{ {\rm ren }}_0 &=-\frac{\partial E^{\rm ren}}{\partial a} \nonumber\\
& =\frac{x^2}{\pi a^2}\int_0^{1} \frac{\sqrt{y^2-y^4}}{e^{2xy}-1}dy - \frac{x^3}{2\pi a^2}\int_0^{1} \frac{y\sqrt{y^2-y^4}}{\text{sinh}^2(xy)}dy.
\label{c9}
\end{align}
Or equivalently, from Eq. \eqref{c5.1},
\begin{align}
F^{ {\rm ren }}_0 =-\frac{x^2}{6\pi a^2}+\frac{x^2}{4a^2}\sum_{k=1}^{\infty}\left\{\left[L_1(2xk)-I_1(2xk)\right]+\left[L_3(2xk)-I_3(2xk)\right]\right\}.
\label{c9.1}
\end{align}
Again, both expressions \eqref{c9} and \eqref{c9.1} are equivalent, however, Eq. \eqref{c9} is more convenient for numerical calculations while Eq. \eqref{c9.1} for investigating the analytical expression for the vacuum force in the regime of small LIV parameter, i.e., $x\gg 1$. In Fig.\ref{fig3} we have plotted the dimensionless vacuum force $\frac{24\pi a^2}{\pi}F^{\rm ren}$, in terms of $x$. One can see that the force shares some very similar aspects with the energy, that is, in the regime where the LIV parameter is very small $x\gg 1$, the force approaches its standard contribution in the absence of Lorentz violation while in the opposite regime $x\ll 1$ it goes to zero. However, just like in the energy case, from the experimental point of view, it does no make sense to talk about values of $x$ smaller than 1 $(\bar{g}>1)$, otherwise the LIV signal would be very strong. Thus, the solid red line is the part of the curve which is realistically acceptable. On the other hand, the blue dashed line is the part of the curve for which $x < 1$.

Now, in order to obtain the limit $\bar{g}\to0$ $(x\gg 1)$, we follow the same procedure used for the Casimir energy, where we adopt the approximation in Eq. \eqref{c6.0}.
\begin{align}
F^{ {\rm ren }}_0 =-\frac{x^2}{6\pi a^2}&+\frac{x^2}{4\pi a^2}\left[\sum_{j=0}^{\infty} \frac{(-1)^{j+1} \Gamma(j+1 / 2)}{\Gamma(3/ 2-j)}\frac{\zeta(2j)}{x^{2j}}\right.\nonumber \\
& \left.+ \sum_{j=0}^{\infty} \frac{(-1)^{j+1} \Gamma(j+1 / 2)}{\Gamma(7/ 2-j)}\frac{\zeta(2j-2)}{x^{2j-2}}\right].
\label{c10}
\end{align}
As a consequence, we are able to write the vacuum force up to order $\bar{g}^3$ as
\begin{align}
F^{\text {ren }}_0=-\frac{\pi}{24a^2}+\frac{\bar{g} \pi^3}{160 a^4}+\frac{5\bar{g}^2 \pi^5}{4032 a^6} +\frac{7\bar{g}^3 \pi^7}{7680 a^8} + \mathcal{O}(\bar{g}^4).
\label{c12}
\end{align}
We can see that the first term in the r.h.s. of the above approximation is the dominant standard contribution of the vacuum force in the absence of LIV \cite{ambjorn1983properties}  while the other terms are corrections due to the presence of a high derivative operator that induces the Lorentz invariance breaking.

\section{Conclusion}
\label{sec6}

In this work, we initiate by projecting the four-dimensional gauge field onto a two-dimensional theory using a higher-derivative order operator. This projection leads to modified equations of motion that rely on the nature of the $n_{\mu}$ Lorentz invariance violation (LIV) parameter. Specifically, we derive space-like and time-like equations of motion, as depicted in Eqs. \eqref{e8.1} and \eqref{e8.2} respectively. Justifying the space-like approach, we demonstrate subluminal behaviour through phase and velocity group analysis (Eq. \eqref{rfr}) and energy behaviour, as depicted in Fig.\ref{fig02}.

By using this dispersion relation, we obtain analytic expressions for the one-dimensional Casimir energy (Eqs. \eqref{c4} and \eqref{c5.1}), which rely on the LIV parameter $\bar{g}$. The behaviour of this energy is illustrated in Fig.\ref{fig1}, displaying convergence to a constant value. However, this closed expression does not allow exploration of the limit where the LIV parameter is small ($\bar{g}\ll1$). To address this, we approximate for large arguments, obtaining an expression for the Casimir energy (Eq. \eqref{c8}) that includes both the usual term for $(1+1)D$ Casimir energy and a correction term. These correction terms depend on the LIV parameter, and in the limit $\bar{g}\to 0$, they vanish entirely, leaving only the conventional term for the one-dimensional Casimir energy.

The Casimir force is also described by the system in Eqs. \eqref{c9} and \eqref{c9.1}, resulting in a closed expression similar to that obtained for the Casimir energy. The behaviour of this analytic expression for the force is depicted in Fig.\ref{fig3}, illustrating the correct interval where the force \eqref{c10} is valid. However, like the Casimir energy \eqref{c5.1}, the analytic expression for the force does not allow exploration of the limit $\bar{g}\ll1$. Consequently, we employ an approximation, leading to Eq. \eqref{c10}, which results in Eq. \eqref{c12}. This equation reveals a repulsive term in the conventional one-dimensional Casimir force, along with correction terms that can be adjusted according to the desired precision. Regardless of the correction order, the force converges negatively as $a$ tends to infinity and diverges positively as $a$ tends to zero.

In conclusion, the combination of the high derivative order operator and dimensional reduction yields analytic expressions for the Casimir energy and Casimir force. Typically, obtaining closed-form expressions in the presence of the LIV parameter requires approximations. However, our approach provides analytic solutions, offering a more comprehensive understanding of the system. Furthermore, the introduction of the LIV parameter allows us to manipulate correction terms to achieve the best fit for both Casimir energy and force, as previously demonstrated. Furthermore, by considering the temperature effect in the present study, it is expected that additional phenomenological details will emerge from the system.


{\acknowledgments}
We would like to thank CNPq, CAPES and CNPq/PRONEX/FAPESQ-PB (Grant nos. 165/2018 and 015/2019), for partial financial support. K.E.L.F would like to thank the Brazilian agency CNPq for financial support. M.A.A, F.A.B and E.P acknowledge support from CNPq (Grant nos. 306398/2021-4,
309092/2022-1, 304290/2020-3). H.F.S.M is partially supported by CNPq under grant no. 311031/2020-0.


\end{document}